\documentclass[10pt,aps,prb,twocolumn,showpacs,floatfix]{revtex4-1}
\usepackage{amsmath,graphics,epsfig,mathrsfs,physics,bbold}
\usepackage{amsfonts}
\usepackage{amssymb}
\usepackage{graphicx}
\usepackage{amsmath}
\usepackage{amsthm}

\usepackage{braket}
\usepackage[english]{babel}
\usepackage[utf8]{inputenc}
\usepackage{graphicx}
\usepackage[colorinlistoftodos]{todonotes}
\usepackage{amsfonts}
\usepackage{empheq}
\usepackage{listings}
\usepackage{hyperref}
\hypersetup{backref=true,pagebackref=true,hyperindex=true,colorlinks=true,breaklinks=true,urlcolor=
red,linkcolor=blue,bookmarks=true,bookmarksopen=true}

\begin{document}

\title{Current-driven Rashba Field in a Magnetic Quantum Well}
\author{Abdulkarim Hariri$^1$}
\author{Meshal Alawein$^2$}
\author{Aurelien Manchon$^{1,2}$}
\email{aurelien.manchon@kaust.edu.sa}
\affiliation{$^1$King Abdullah University of Science and Technology (KAUST), Physical Science and Engineering Division (PSE), Thuwal 23955-6900, Saudi Arabia.}
\affiliation{$^2$King Abdullah University of Science and Technology (KAUST), Computer, Electrical and Mathematical Science and Engineering Division (CEMSE), Thuwal 23955-6900, Saudi Arabia.}

\begin{abstract}
In materials lacking inversion symmetry, the spin-orbit coupling enables the direct connection between the electron's spin and its linear momentum, a phenomenon called inverse spin galvanic effect. In magnetic materials, this effect promotes current-driven torques that can be used to control the magnetization direction electrically. In this work, we investigate the current-driven inverse spin galvanic effect in a quantum well consisting in a magnetic material embedded between dissimilar insulators. 
Assuming the presence of Rashba spin-orbit coupling at the interfaces, we investigate the nature of the non-equilibrium spin density and the influence of the quantum well parameters. We find that the torque is governed by the interplay between the number of states participating to the transport and their spin chirality, the penetration of the wave function into the tunnel barriers, and the strength of the Rashba term.   
\end{abstract}
\maketitle


\section{Introduction}
Spin transfer torque\cite{slonczewski1996current,berger1996emission} has been the primary method to record information electrically in small magnetic devices, leading to the successful development of modern Magnetic Random Access Memories \cite{kent2015new}. This effect consists in transferring spin angular momentum from a spin-polarized electrical current, obtained by passing the current through a reference ferromagnet, into an adjacent magnetic layer, thereby exerting a torque on its magnetization. An alternative method to generate such a torque is to exploit the spin-orbit interaction of magnetic materials or heterostructures lacking inversion symmetry \cite{miron2010current,liu2012spin}. In these systems, a metal with strong spin-orbit interaction acts as a source of non-equilibrium spin density that can in turn torque the magnetization of the free adjacent magnetic layer, a phenomenon called spin-orbit torque \cite{manchon2018current}. This torque, which does not require a reference ferromagnet, enables current-driven magnetization switching \cite{miron2010current,Miron2011b,liu2012spin} and excitations \cite{liu2012current,demidov2012magnetic}, as well as fast magnetic domain wall motion \cite{Miron2011,emori2013current,ryu2013chiral}.\par

Two main classes of physical mechanisms have been identified as a source of spin-orbit torque: the spin Hall effect \cite{Dyakonov1971b,Hirsch1999}, emerging from adjacent heavy metals, and the inverse spin galvanic effect \cite{Ivchenko1978,Edelstein1990}, emerging either in bulk non-centrosymmetric magnets\cite{Garate2009,Bernevig2005c} [strained (Ga,Mn)As\cite{Chernyshov2009}, NiMnSb \cite{Ciccarelli2016}, Mn$_2$Au\cite{Meinert2018,Bodnar2018} or CuMnAs\cite{Wadley2016}] or at interfaces between a magnet and a heavy metal\cite{manchon2008theory}. While inverse spin galvanic effect is the sole mechanism occurring in bulk non-centrosymmetric systems, magnetic multilayers are expected to exhibit both spin Hall and inverse spin galvanic effects, together with several additional effects such as spin swapping \cite{Saidaoui2016} and spin precession around the interfacial spin-orbit field \cite{Amin2018}. It is therefore uneasy to disentangle all these effects. Interestingly, recent studies have managed to get rid of the spin Hall effect by considering the interface between a ferromagnet and an insulator. For instance Emori et al. \cite{Emori2016} recently reported spin-orbit torque in Ti/NiFe/AlOx trilayer. The torque observed by the authors has the characteristic of a magnetic field, which constitutes a signature of inverse spin galvanic effect \cite{manchon2008theory}. In another inspiring experiment, Qiu et al. \cite{Qiu2015} demonstrated that the sign of the spin-orbit torque can be completely reversed by tuning the oxidation of Pt/CoFeB/MgO multilayers, suggesting that the spin-orbit torque at CoFeB/MgO interface can dominate over the torque arising from Pt.\par

Inspired by these results, the present work investigates the inverse spin galvanic effect in a quantum well composed of a ferromagnet embedded between dissimilar insulators. Our research intends to determine the nature of spin-orbit torque in such a structure, the role of quantum confinement, and the interplay between the spin-orbit coupling at opposite interfaces. Solving Schr\"odinger equation for a quantum well with interfacial Rashba spin-orbit coupling, we find that the torque is governed by the interplay between the number of states participating to the transport and their spin chirality, the penetration of the wave function into the tunnel barriers, and the strength of the Rashba term. 
\begin{figure}[h!]
\includegraphics[width=8cm]{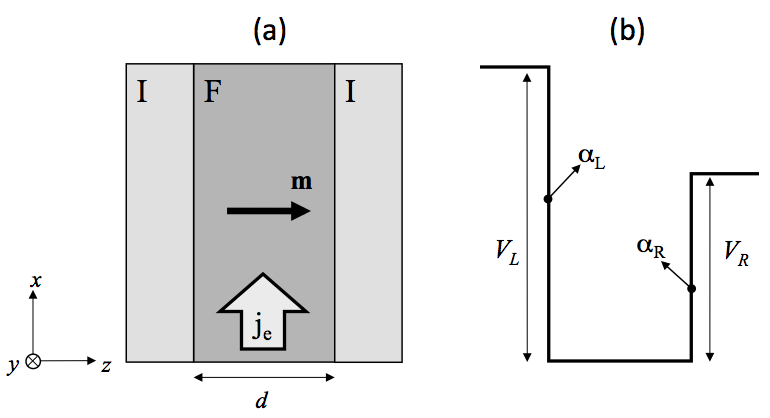}
\caption{(a) Schematics of the I/F/I trilayer. The magnetization (black arrow) is set perpendicular to the plane, while the current (thick arrow) is injected along the $x$ direction. (b) Corresponding potential profile of the quantum well. $V_{L,R}$ are the potential of the left and right insulators, respectively, and $\alpha_{L,R}$ are the Rashba parameters at these interfaces.\label{fig:Fig0}}
\end{figure}  

\section{Model}

The magnetic quantum well is depicted in Fig. \ref{fig:Fig0} and consists of a ferromagnetic layer (F), with magnetization fixed along {\bf z}, normal to the planes, embedded between two non-magnetic insulators (I). The Hamiltonian of the system is given by
\begin{align}
&\hat{H} = \hat{H}_0 + \hat{H}_{R},\\
&{{\hat H}_0} = \left[ {\frac{{{{\hat p}^2}}}{{2m}} + V\left( z \right)} \right]\hat I+\Delta\boldsymbol{\hat\sigma}\cdot{\bf m},\\
&{{\hat H}_R}= {\alpha _{L}}\left( {{{\mathbf{z}}} \times {\mathbf{p}}} \right) \cdot {\boldsymbol{\hat \sigma }}\delta\left( z \right) - {\alpha _{R}}\left( {{{\mathbf{z}}} \times {\mathbf{p}}} \right) \cdot {\boldsymbol{\hat \sigma }}\delta\left( {z - d} \right) 
\end{align} 
where $\hat{H}_0$ is the free electron Hamiltonian with potential $V(z)$ such that
\begin{equation}
V\left( z \right) = \left\{ \begin{gathered}
 {V_L},\;\;z < 0 \hfill \\
 0,\;\;\;\;\;0 \le z \le d \hfill \\
 {V_R},\;\;z > d \hfill \\ 
\end{gathered} \right. \hfill \\
\end{equation}
$V_{L,R}$ are the left and right potential barrier heights, $\Delta$ is the exchange energy, $\hat{I}$ is the identity matrix in spin space and $\hat{\bm\sigma}$ is the vector of Pauli spin matrices. $\hat{H}_R$ describes the interfacial spin-orbit coupling, where $\alpha_{L}$, $\alpha_R$ are the Rashba parameters at the left and right interfaces, and $\delta(z)$ is Dirac delta function.\\

The wavefunction $\psi(z)$ is obtained by matching the generic solutions of Schr\"odinger's equation in each region. Due to the interfacial Rashba potential, $\frac{d \psi}{d z}$ is discontinuous at the interfaces, giving a boundary condition of the form (e.g., at the right interface)
\begin{equation}
{\left. {\frac{{d\psi }}{{dz}}} \right|_{z = {d^ + }}} - {\left. {\frac{{d\psi }}{{dz}}} \right|_{z = {d^ - }}} = - \frac{2m\alpha_{R}}{\hbar}\left( \mathbf{z} \times \mathbf{k} \right) \cdot \hat{\boldsymbol{\sigma }}\psi(d)
\end{equation}
Therefore, in order to obtain the electronic states, the quantization rule is solved numerically, which gives the eigenmomenta corresponding to the bound states of the well. To analyze the system's linear response to an external electrical field, we use Kubo formula within the relaxation time approximation. We emphasize that linear response theory normally results in contributions to non-equilibrium properties arising from intraband and interband transitions \cite{li2015intraband,Freimuth2014a}. The former is responsible for the inverse spin galvanic effect, and thereby a field-like torque, while the latter produces a correction that gives rise to a so-called damping-like torque. This correction to inverse spin galvanic effect comes from the precession of the non-equilibrium spin density around the magnetic exchange. In the present case, since the Rashba interaction is interfacial, we expect this precession to be small and neglect the interband contributions. Therefore, the current-driven spin density reads
\begin{equation}
\small S_y(z) = -\frac{\tau e E}{2\pi} \int \frac{d^2\bf{k}}{4\pi^2}\sum_{n,s} \langle n,s |\hat{v}_x| n,s \rangle \langle n,s |\hat\sigma_y| n,s \rangle \delta(\epsilon_{\bf F} - \epsilon_{\bf k}^{n,s}),
\end{equation}
where $\tau$ is the relaxation time, $\epsilon_{\rm F}$ is Fermi energy, $E$ is the electric field applied along the $x$-direction, and $| n,s \rangle$ is the eigenstate $n$ with spin $s$, momentum ${\bf k}$ and eigenenergy $\epsilon_{\bf k}^{n,s}$.

\section{Results}

The nature of current-driven inverse spin galvanic effect in a magnetic quantum well depends on the chirality of the confined states and their respective contribution to the spin density. Therefore, one needs to first understand how Rashba spin-orbit coupling influences the confinement itself and how the various states contribute to the non-equilibrium transport. After that, we inspect how the well's geometry (width and height) influences Rashba spin splitting and hence the spin transport. 

\begin{figure}
    \includegraphics[height=2.8in]{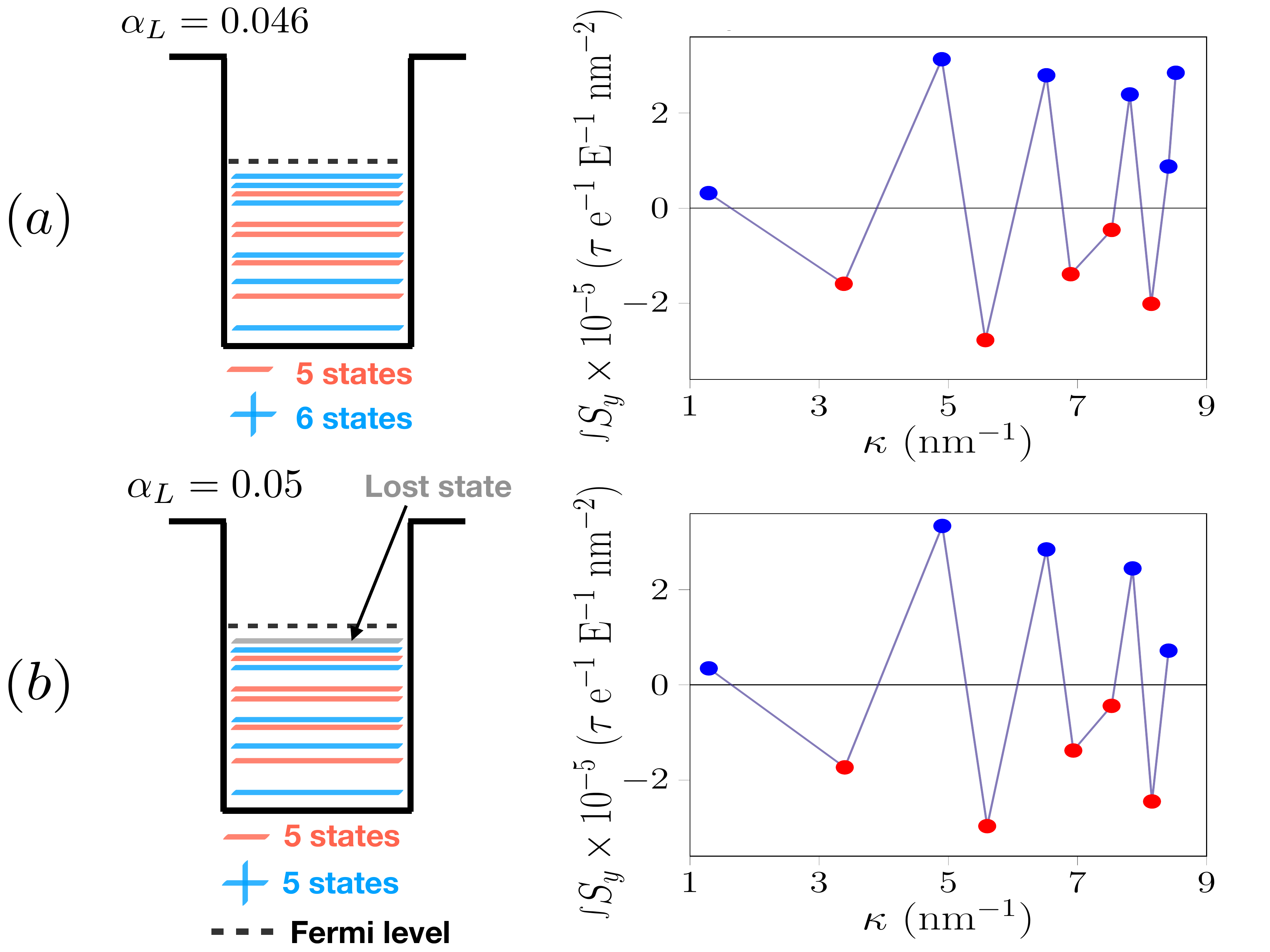}
   \caption{(Color online) Quantum well states (left panels) and their contributions to the non-equilibrium spin density (right panels). (a) The interfacial Rashba parameters are set to $\alpha_{R} = 0$ and $\alpha_{L}$ = 0.046 eV$\cdot$nm$^2$, 11 states are present below Fermi level. (b) The interfacial Rashba parameters are set to $\alpha_{R} = 0$ and $\alpha_{L}$ = 0.05 eV$\cdot$nm$^2$, one state with positive chirality is lost. The blue and red colors correspond to positive and negative contributions to the non-equilibrium spin density, respectively.\label{fig:fig2}}
  \end{figure}
\subsection{Quantum confinement of Rashba states}
We first consider a symmetric quantum well ($V_L = V_R = $ 5 eV) with a width $d$=2.2 nm, a Fermi energy of 3 eV, and the exchange energy $\Delta$ = $0.25$ eV. At metallic interfaces, the value of Rashba parameter lies typically in the range $\tilde{\alpha}_{L,R} \approx 10^{-2} - 10^{-1}$ eV$\cdot$nm \cite{LaShell1996,Ast2007,King2011,Grytsyuk2016}. In the absence of further estimate at I/F interfaces, we set the interfacial Rashba parameter to $\alpha_{L,R} = \tilde{\alpha}_{L,R} a_0\approx10^{-2} - 10^{-1}$ eV$\cdot$nm$^2$, where $a_0\approx 0.3$ nm is the typical distance over which the interfacial potential drops. This simple model allows us to clearly identify the influence of the Rashba parameter on the bound states of the well, as well as on their overall contribution to the current-driven inverse spin galvanic effect. Each bound state is defined by its spin chirality. We use the convention that $+$ ($-$) spin chirality designates those states contributing positively (negatively) to the total spin density. An example is given in Fig. \ref{fig:fig2}(a). There, the interfacial Rashba parameter is set to $\alpha_{R}=0$, and $\alpha_{L}=0.046$ eV$\cdot$nm$^2$. On the left panel, the 11 bound states located below Fermi level are schematically depicted, while on the right panel the contribution of each state to the total non-equilibrium spin density $\int_0^d S_y dz$ as a function of the quantized wave vector ${\kappa}$ is reported. In this particular example, 6 states contribute to a positive spin density (blue), while 5 states contribute to a negative spin density (red). Interestingly, all states contribute to the non-equilibrium spin density, except the lowest energy state. When the Rashba parameter of the left interface is increased to $\alpha_{L}=0.05$ eV$\cdot$nm$^2$ [see Fig. \ref{fig:fig2}], the state closest to Fermi level is lost, which reduces the positive contribution to the spin density. 

\begin{figure}
    \includegraphics[height=3.8in]{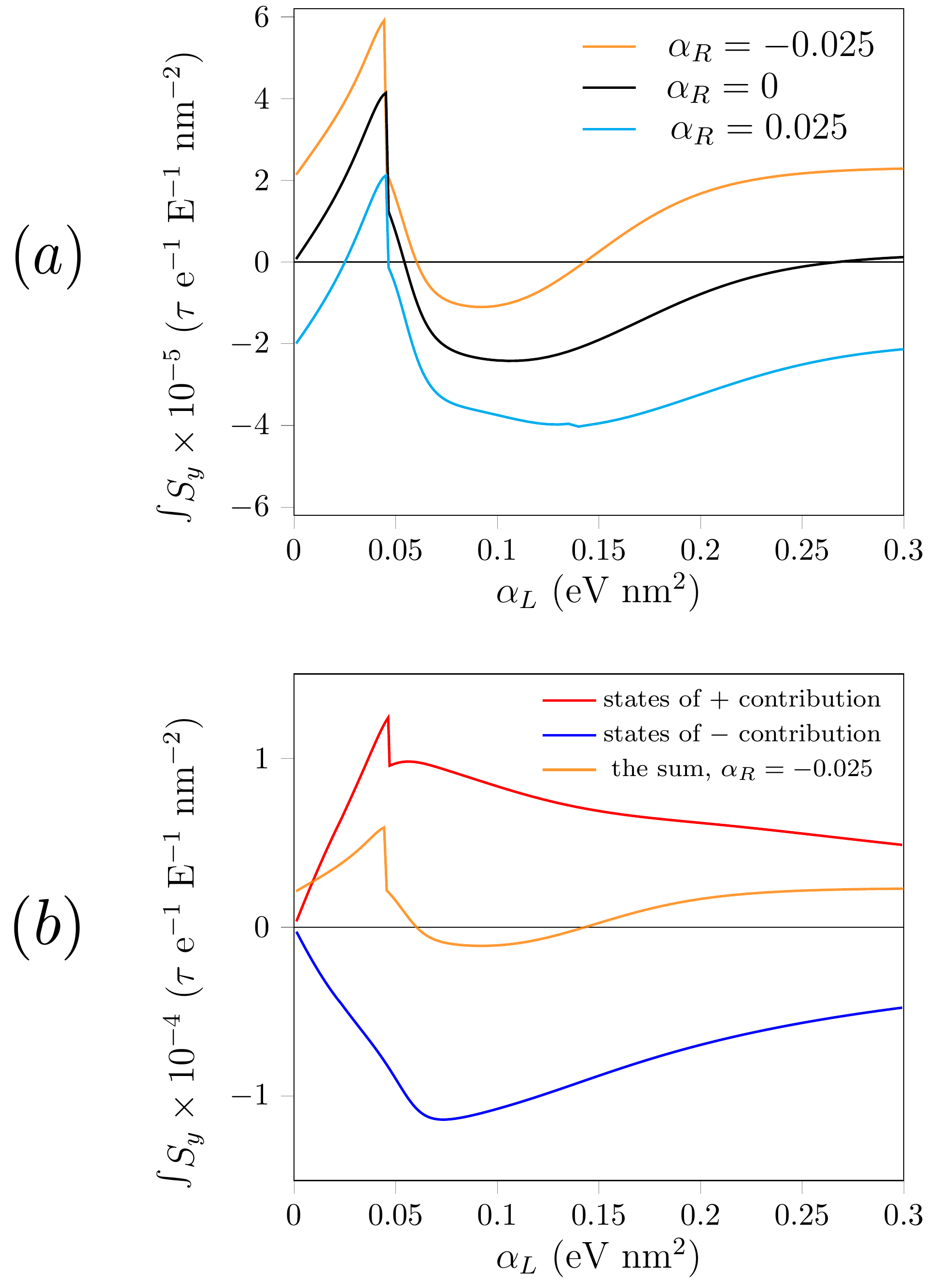}
   \caption{(Color online) (a) Total spin density dependence on Rashba parameter (at the left interface) is plotted for three different values of $\alpha_{R}$. Changing $\alpha_{R}$ merely shifts the curve vertically. (b) Spin chirality-resolved contributions to the total density for  $\alpha_{R}=-0.025$ eV$\cdot$nm$^2$. \label{fig:fig3}}
  \end{figure}

The change of the number of states upon modifying the interfacial Rashba parameter has a dramatic impact on the overall magnitude of the current-driven spin density. This can be seen in Fig. \ref{fig:fig3}(a), where the non-equilibrium spin density is represented as a function of $\alpha_{L}$ for three distinct values of $\alpha_{R}$. This graph shows three regions.
 
For 0$<\alpha_{L}<$0.049 eV$\cdot$nm$^2$, the non-equilibrium spin density increases linearly with $\alpha_{L}$, as expected from the simple Rashba model. In this range, the interfacial Rashba potential remains much smaller than the barrier height, so that the number of states contributing to the transport remains fixed and is equal to 11, as in Fig \ref{fig:fig2}(a). A jump occurs at $\alpha_{L}$=0.049 eV$\cdot$nm$^2$, due to the loss of one confined state, as explained in Fig. \ref{fig:fig2}. A finer analysis can be obtained from Fig. \ref{fig:fig3}(b), where the non-equilibrium spin density arising from the contribution of states with $+$ and $-$ chiralitites is reported. As explained earlier, Rashba parameter changes the potential the electrons feel at the interface, which in return affects the number of allowed bound states. The drop shown in Fig. \ref{fig:fig3}(b) is attributed to the loss of a state with $+$ chirality [shown in Fig. \ref{fig:fig3}(a) as well]. 
  
By further increasing the Rashba parameter, 0.049$<\alpha_{L}<$0.1 eV$\cdot$nm$^2$, Fig. \ref{fig:fig3}(a) also shows a decrease in the total spin density, due to the $-$ states progressively overcoming the $+$ states, while keeping the number of confined states constant [see Fig. \ref{fig:fig3}(b)]. In this region, increasing the Rashba parameter results in a chirality-dependent enhancement of the confining potential. In other words, the ability of the wave function to penetrate the barrier reduces and results in a progressive quenching of the current-driven spin density. Since this quenching is chirality-dependent, the overall slope is non-linear for $\alpha_L>$0.1 eV$\cdot$nm$^2$.

\begin{figure}[h!]
    \includegraphics[height=3.8in]{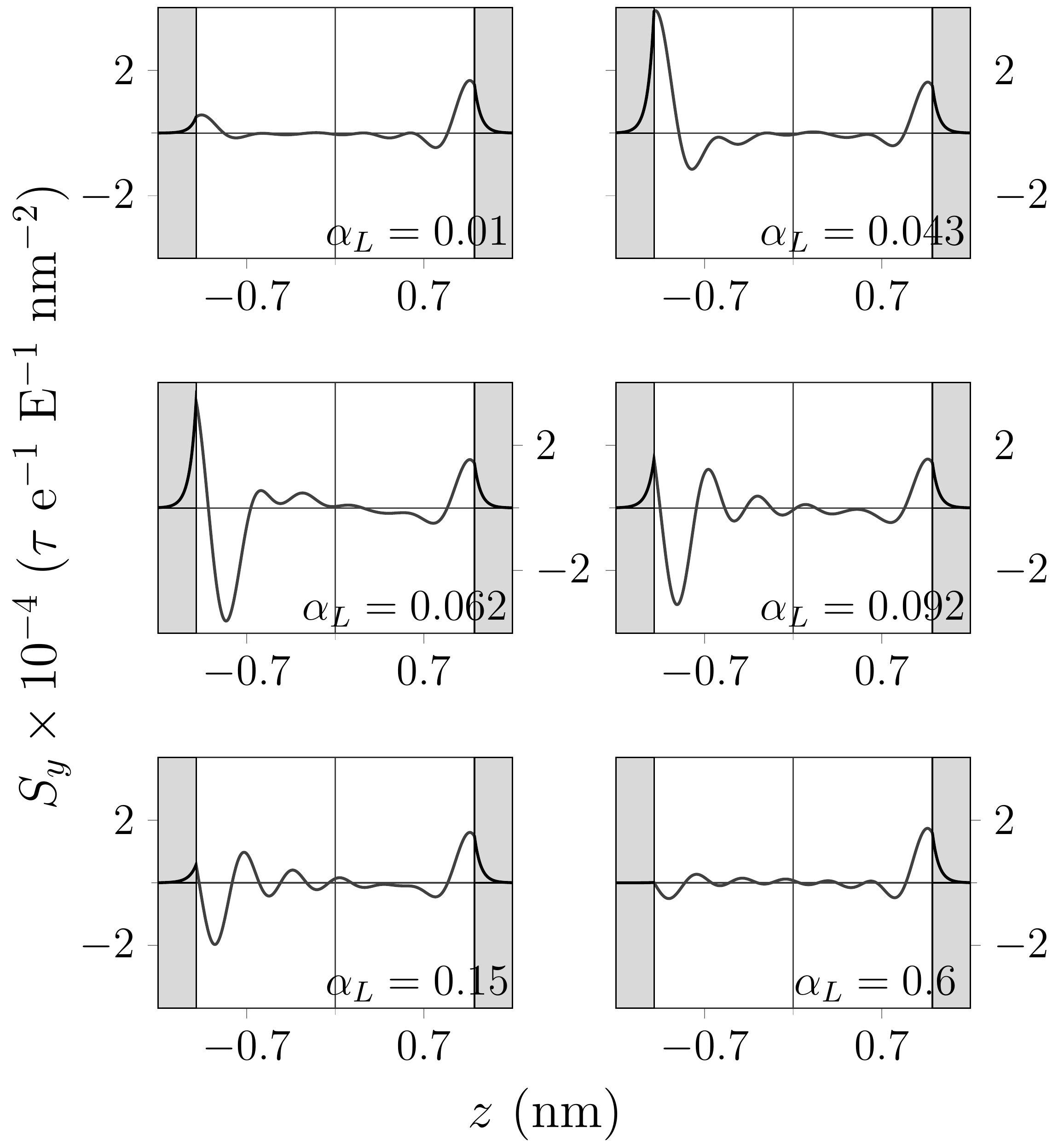}
   \caption{Non-equilibrium spin density profile across the magnetic quantum well for various values of the left interfacial Rashba parameter $\alpha_{L}$, while keeping $\alpha_{R}= - 0.025$ eV$\cdot$nm$^2$.\label{fig:fig5}}
  \end{figure}
  
This qualitative picture is confirmed by inspecting the non-equilibrium spin density profile across the magnetic quantum well, displayed in Fig. \ref{fig:fig5}. Upon increasing $\alpha_{L}$, the value of the spin density at the left interface increases progressively, until $\alpha_{L} = 0.062$ eV$\cdot$nm$^2$. After that, the penetration of the spin density into the left tunnel barrier is substantially quenched, resulting in a continuous decrease in the overall non-equilibrium spin density at the left interface.

\subsection{Influence of the thickness}

\begin{figure}[h!]
    \includegraphics[height=2.9in]{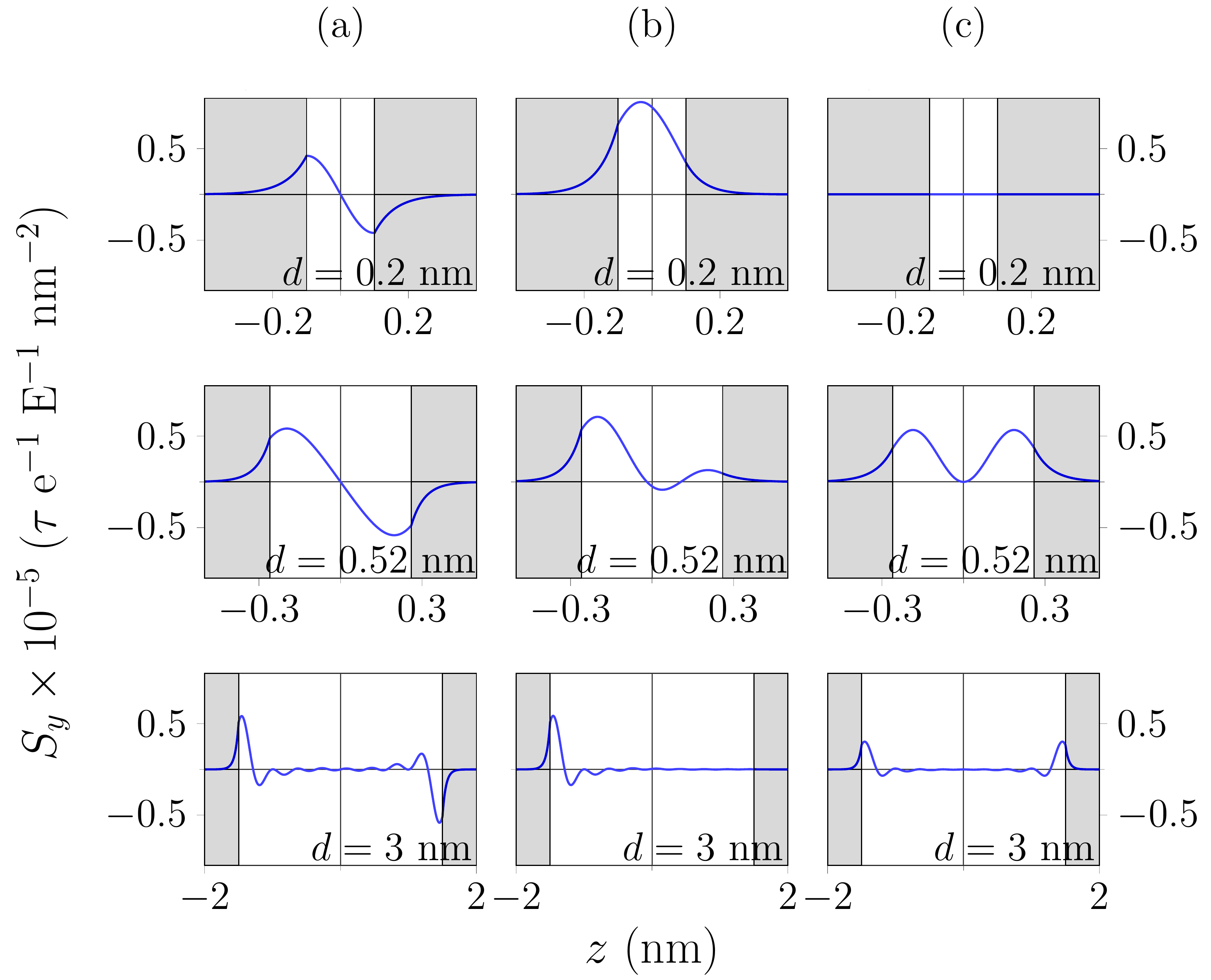}
   \caption{(Color online) Non-equilibrium spin density profile across the magnetic quantum well for three different thicknesses and in (a) $\alpha_{L} = \alpha_{R}=0.01$  eV$\cdot$nm$^2$, (b) $\alpha_{L} = 0.01$ eV$\cdot$nm$^2$, $\alpha_{R}=0$, and (c) $\alpha_{L} = - \alpha_{R}=0.01$ eV$\cdot$nm$^2$. \label{fig:fig6}}
  \end{figure}

An important question is whether the two interfaces can be considered as acting independently. Qualitatively, one expects that for narrow quantum wells, the two interfaces "talk" to each other, producing a non-equilibrium spin density that is not the sum of the independent interfaces, while at large thicknesses, the total spin density is simply the sum of the interfacial spin densities. To validate this idea, we now consider the non-equilibrium spin density profile for three complementary cases, displayed in Fig. \ref{fig:fig6}: (a) $\alpha_{L}= \alpha_{R}$, (b) $\alpha_{L} \neq0$, $\alpha_{R}=0$, and (c) $\alpha_{L} = - \alpha_{R}$. \par

\begin{figure}[h!]
    \includegraphics[height=2.5in]{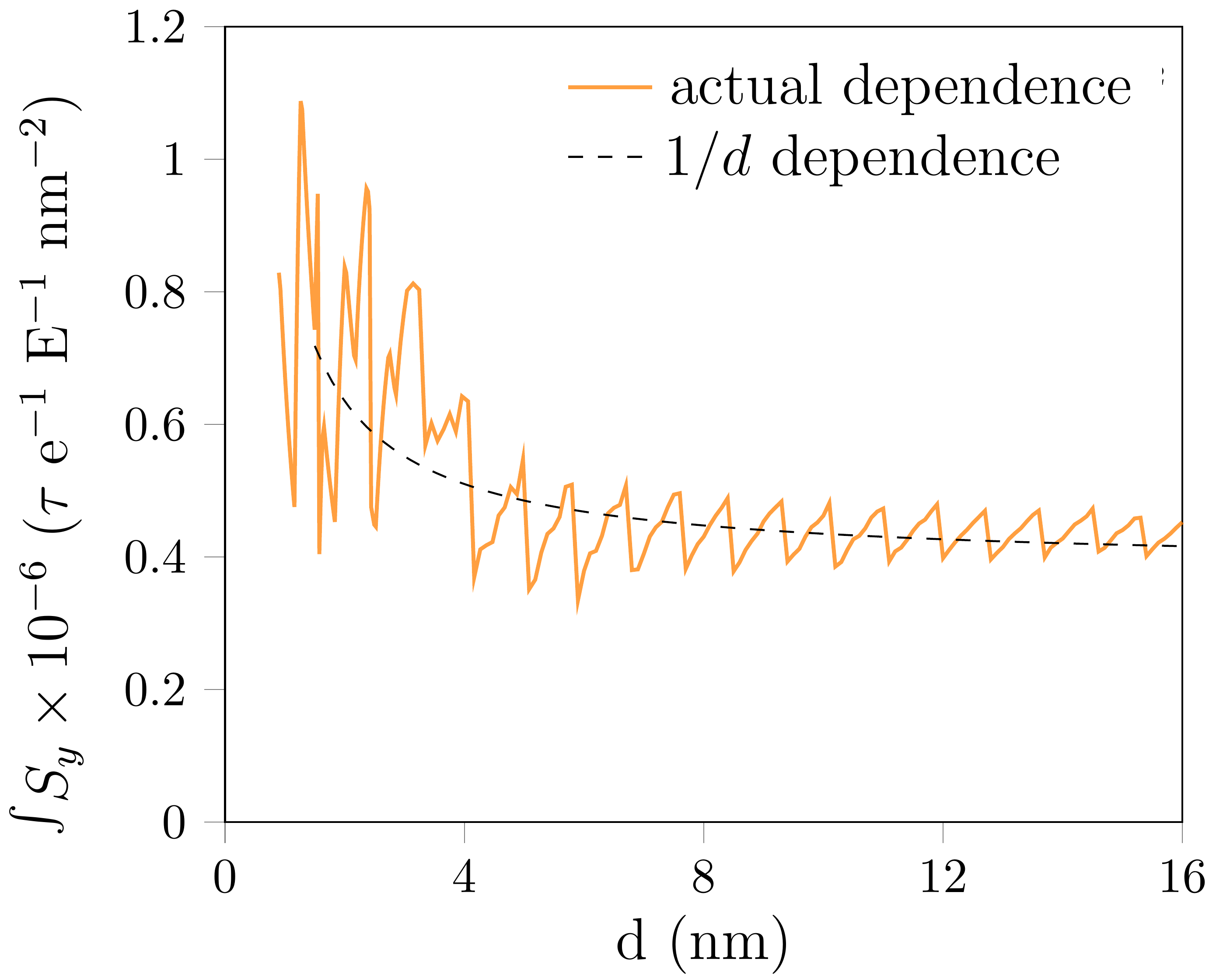}
   \caption{(Color online) Total non-equilibrium spin density as a function of the well's width $d$ with parameters set to $\alpha_{L} = - \alpha_{R} = 0.001$ eV$\cdot$nm$^2$. The dashed line is a guide for the eye and scales with $1/d$.\label{fig:fig7}} \end{figure}

In the large thickness case, $d=3$ nm (bottom panels), the spin density is well localized at each interface such that the contribution of the two interfaces are mostly independent of each other. Notice though that the overall magnitude of the spin density in the case (a) $\alpha_{L} = \alpha_{R}$ is smaller than for (a) $\alpha_{L}=-\alpha_{R}$, due to the different confining potential as discussed in the previous section. More specifically, case (a) has twice more states below Fermi level than case (c) (27 in the former and 14 in the latter), showing that the impact of the spin-orbit interaction on the confinement can be quite dramatic, even for $d$=3 nm. Upon decreasing the thickness, the wave function progressively envelopes both interfaces, resulting in a more complex thickness dependence in this parameter region: in case (a), the spin density slightly decreases due to the loss of states, while in case (b), the spin density increases. In case (c), the spin density passes by a maximum at $d = 0.52 $ nm and completely vanishes at $d = 0.2 $ nm, due to the expulsion of the last confined state. The overall thickness dependence of the total spin density is reported in Fig \ref{fig:fig7}. At large thicknesses, the spin density displays the expected $1/d$-dependence, modulated by the jump in the number of states accommodated by the quantum well. At small thicknesses, below 4 nm, the oscillation becomes hectic, revealing the interplay between the two interfaces.

\linespread{0}
\section{Conclusion}

Spin-orbit torques have been mostly investigated in magnetic multilayers involving a ferromagnet adjacent to a heavy metal (W, Ta, Pt etc.), the latter providing the strong spin-orbit coupling necessary to sustain large spin Hall effect and inverse spin galvanic effect \cite{manchon2018current}. In recent years, the emergence of spin-orbit torque at the interface between the ferromagnet and an oxide has attracted increasing attention \cite{emori2016interfacial,Qiu2015}. Inspired by these experiments, we investigated the nature of non-equilibrium spin density in a magnetic quantum well, with interfacial Rashba interactions. Rashba interaction influences the inverse spin galvanic effect at three levels: First, it induces spin-momentum locking and is therefore directly responsible for the non-equilibrium spin density. Second, it governs the penetration of the wave function inside the tunnel barrier. Third, it controls the quantum confinement and thereby the number of states participating to the transport. The interplay between these three effects results in a complex, non-linear dependence of the non-equilibrium spin density as a function of Rashba parameter, and in an unconventional thickness dependence at small thicknesses.\par

We conclude by commenting on the experimental relevance of the present study. In experiments, insulators are usually oxidized metals such as MgO or AlOx. A series of experiments have shown the dramatic influence of interfacial oxidation on perpendicular magnetic anisotropy \cite{Manchon2008c,Ikeda2010} and spin-orbit torque \cite{Qiu2015,Demasius2016,An2018,Zhang2019}. First principle calculations have confirmed that the hybridization of O 2$p$ orbitals with the 3$d$ orbitals of the adjacent ferromagnet leads to complex interfacial physics, including the modification of the interfacial spin polarization \cite{Oleinik2000,Belashchenko2004}, enhanced perpendicular magnetic anisotropy \cite{Yang2011}, and Dzyaloshinskii-Moriya interaction \cite{Belabbes2016b}. To date, the influence of interfacial oxidation on the Rashba effect remains unaddressed. Although the present study does not cover the complex interfacial orbital hybridization, it emphasizes that optimal spin-orbit torque could be achieved by reaching a compromise between barrier height (controlling the wave function penetration) and interfacial potential gradient (controlling the Rashba effect).


 
\section*{Acknowledgement}
 This work was supported by the King Abdullah University of Science and Technology (KAUST).

\bibliographystyle{unsrt}
\bibliography{mybib1}

\end{document}